\documentclass[fleqn,12pt,twoside]{article}
\usepackage[headings]{espcrc1}

\readRCS
$Id: espcrc1.tex,v 1.2 2004/02/24 11:22:11 spepping Exp $
\usepackage{graphicx}
\usepackage[figuresright]{rotating}

% ...

\newcommand{\AmS}{{\protect\the\textfont2
 A\kern-.1667em\lower.5ex\hbox{M}\kern-.125emS}}

\newcommand{\jpsi}{{\rm J}/\psi}

\title{PHENIX results on $\jpsi$ production in Au+Au and Cu+Cu collisions at $\sqrt{s_{NN}}=200$~GeV}

\author{H. Pereira Da Costa\address[CEA]{DAPNIA, CEA Saclay, Gif/Yvette, France},
for the PHENIX Collaboration\footnote{for the full list of PHENIX authors and acknowledgements,
see Appendix ``Collaboration'' of this volume.
}}

\runtitle{PHENIX results on $\jpsi$ production in Au+Au and Cu+Cu}
\runauthor{H. Pereira Da Costa for the PHENIX Collaboration}

\begin{document}
\maketitle

\begin{abstract}
Heavy quarkonia production is predicted to be sensitive to the formation of the quark gluon plasma in relativistic heavy ion collisions via competing mechanisms such as color screening and/or quark recombination. During 2004 and 2005 RHIC data taking periods, the PHENIX collaboration has measured $\jpsi$ decay into lepton pairs both at mid and forward rapidity in Au+Au and Cu+Cu collisions at $\sqrt{s_{NN}}=200$~GeV. We present the $\jpsi$ nuclear modification factor and the $\jpsi$ mean square transverse momentum as a function of the collision centrality for both systems, as well as the rapidity dependence of the $\jpsi$ yield for different centrality classes. It is compared to different theoretical predictions. All Au+Au and Cu+Cu results shown here are preliminary.
\end{abstract}

% main text
\section{Introduction}
It has been established at lower energies that the $\jpsi$s created in nuclear collisions are suppressed when the system size and the centrality of the collision increase~\cite{SPS}. For heavy enough nuclei the suppression exceeds the {\em normal} nuclear absorption observed with light nuclei. Several mechanisms, including color screening in the quark gluon plasma, have been proposed to explain this {\em abnormal} suppression observed at low energy. Measurements performed at higher energy are therefore crucial to constrain the models and discriminate between them. Moreover they may uncover new mechanisms such as quark recombination.

\section{Experimental setup and datasets}
The PHENIX experiment measures $\jpsi\rightarrow e^+e^-$ decay at mid rapidity ($|\eta|<0.35$) and $\jpsi\rightarrow \mu^+\mu^-$ decay at forward rapidity ($|\eta|\in[1.2,2.2]$). Electrons are identified using RICH detectors and by matching their momentum measured in drift chambers with the energy deposited in electromagnetic calorimeters. Muons are selected using a thick absorber located close to the interaction point, tracked using cathode strip chambers and triggered on using a succession of Iarocci tube planes and steel walls. 

The results presented here correspond to 240~$\mu$b$^{-1}$ Au+Au collisions collected in 2004, and 3.1 nb$^{-1}$ Cu+Cu collisions in 2005, both at $\sqrt{s_{NN}}=200$~GeV. For each centrality, rapidity or transverse momentum bin, the $\jpsi$ yield is obtained by counting the number of unlike-sign dilepton pairs in a mass window centered on the $\jpsi$ mass after the background has been subtracted using either an event mixing technique or the mass distribution of the like-sign pairs. It is corrected by the efficiency and acceptance of the spectrometer as well as the trigger efficiency and normalized to the number of recorded collisions. This yield enters the numerator of the nuclear modification factor which is the ratio between the measured yield in A+B collisions over the expected yield when assuming binary scaling from p+p collisions. The $\jpsi$ yield measured in p+p collisions is taken from PHENIX 2003 published results~\cite{PHENIX_RUN3}. Systematic errors have been assigned to each contribution. For Au+Au collisions, the dominant systematic error comes from the background subtraction, due to the poor signal over background ratio, especially for central collisions. 

\section{Results}
\begin{figure}[h]
\vspace*{-5mm}
\begin{center}
\includegraphics[width=75mm,clip=true]{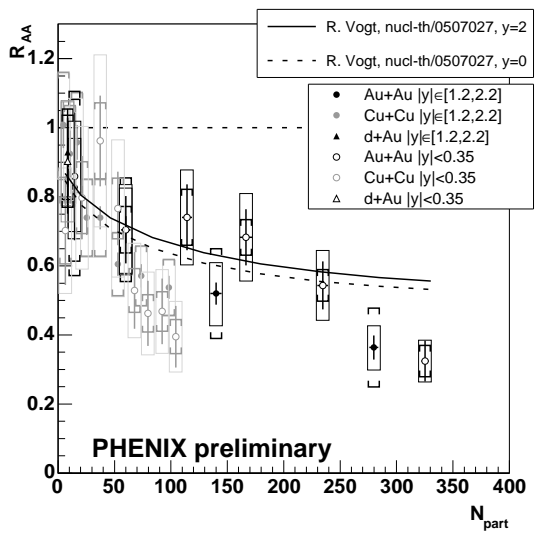}
\end{center}
\vspace*{-15mm}
\caption{\label{fig_vogt}$\jpsi$ nuclear modification factor as a function of the number of participant, compared to R. Vogt predictions for the normal nuclear absorption. Vertical bars are statistical errors, brackets are point to point systematics and boxes are global systematics.}
\vspace*{-5mm}
\end{figure}

\begin{figure}[h]
\vspace*{-5mm}
\begin{center}
\begin{tabular}{cc}
\includegraphics[width=75mm,clip=true]{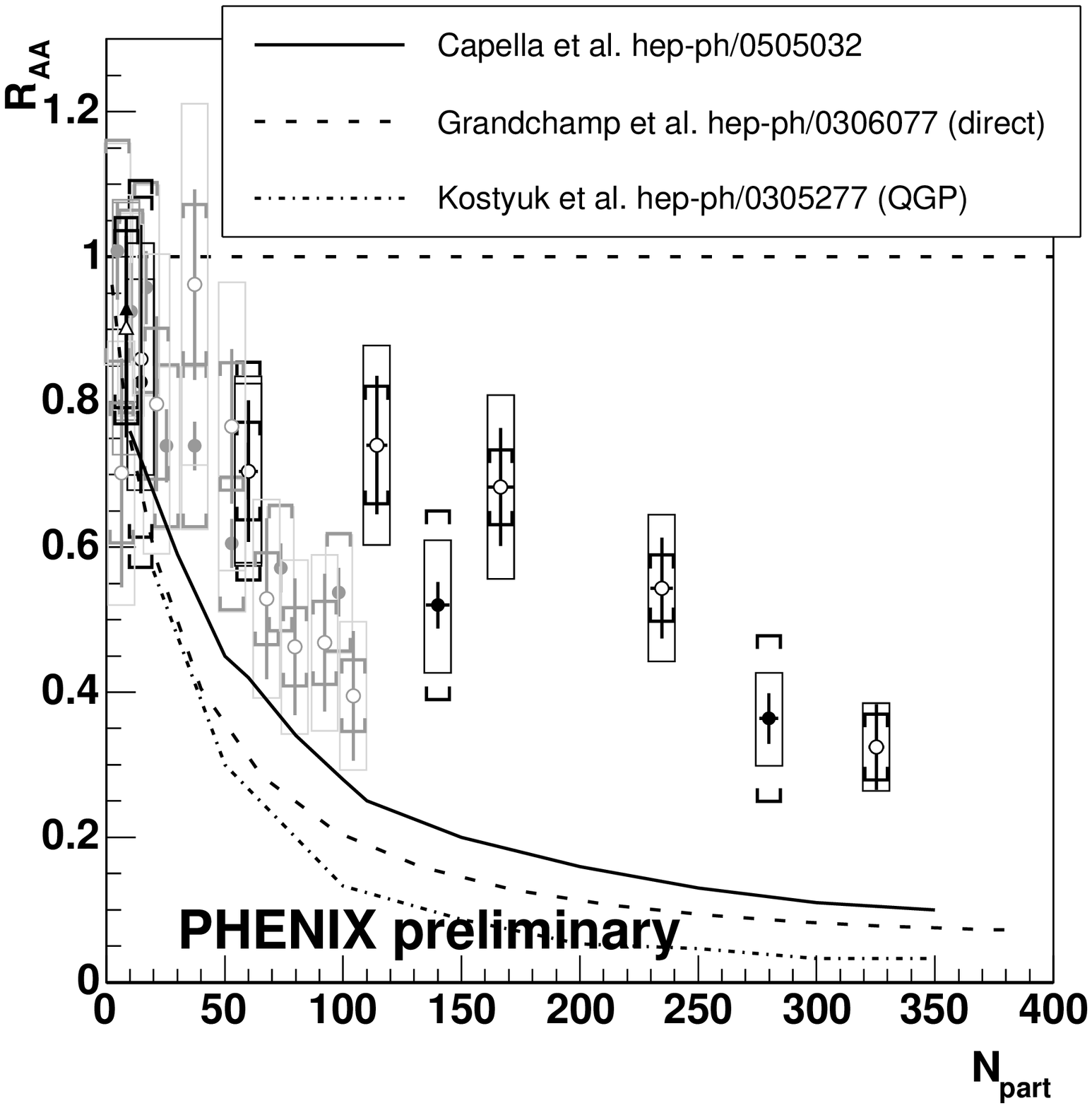}&
\includegraphics[width=75mm,clip=true]{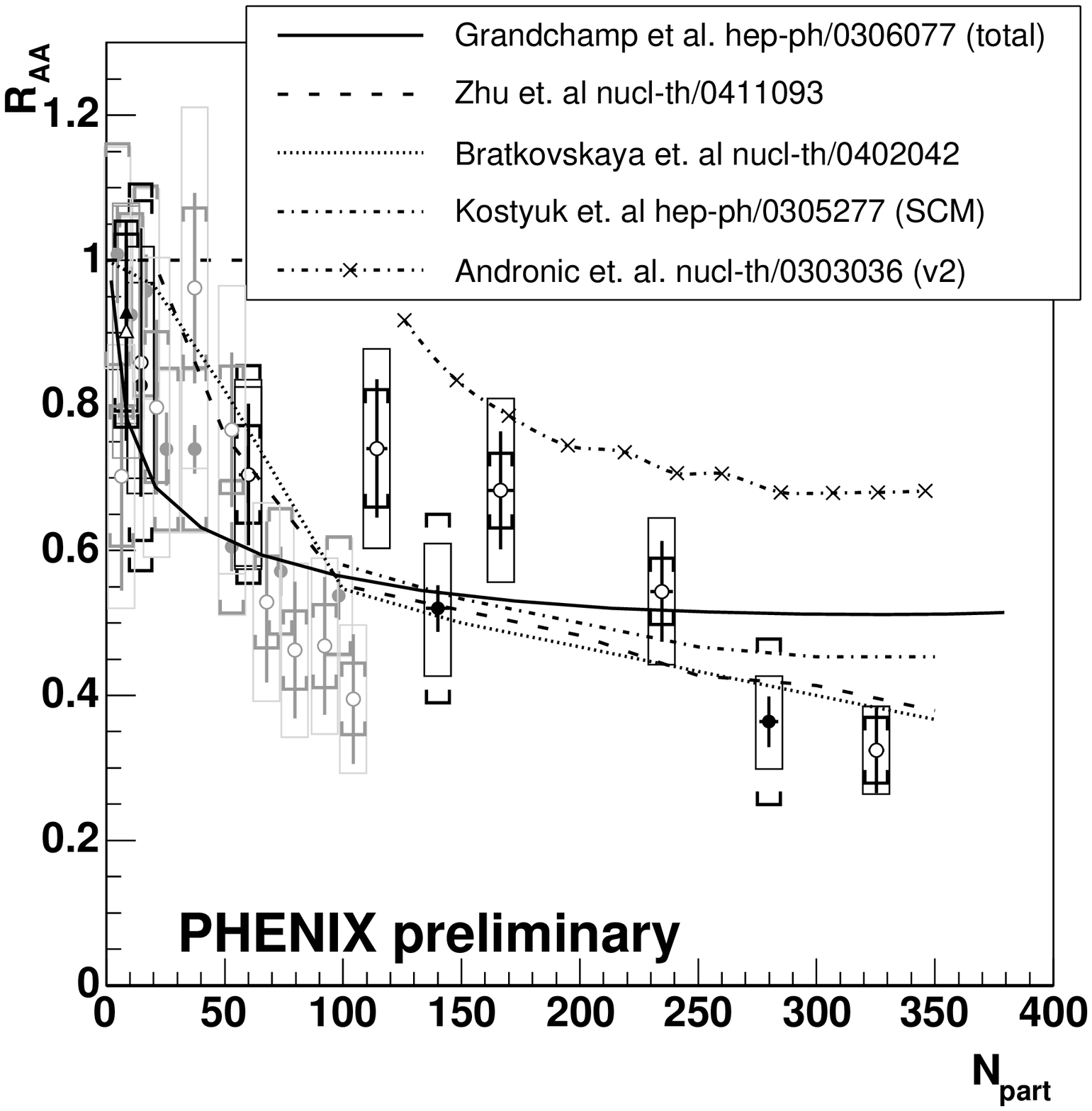}\\
\end{tabular}
\end{center}
\vspace*{-15mm}
\caption{\label{fig_theo}$\jpsi$ nuclear modification factor as a function of the number of participant compared to various models of final state interaction in the medium.}
\vspace*{-5mm}
\end{figure}

Figures~\ref{fig_vogt} and~\ref{fig_theo} represent the $\jpsi$ nuclear modification factor as a function of the number of participant for Au+Au and Cu+Cu collisions both at mid an forward rapidity together with the published measurement from d+Au~\cite{PHENIX_RUN3}. A suppression of about a factor 3 is observed for the most central collisions. 

On figure~\ref{fig_vogt}, the data are compared to R. Vogt prediction~\cite{VOGT}, assuming 3~mb nuclear absorption on top of EKS98 shadowing. The prediction miss the most central points. Moreover, from~\cite{PHENIX_RUN3} it appears that 3~mb is an upper bound for the nuclear absorption cross-section in d+Au collisions. 

On figure~\ref{fig_theo} the same data are compared to two sets of models. They all involve additional final state interactions and reproduce both the CERN SPS results~\cite{SPS} and the PHENIX low statistics results from 2002~\cite{PHENIX_RUN2}. The predictions plotted on the left panel (\cite{CAPELLA,GRANDCHAMP,KOSTYUK}) overestimate the $\jpsi$ suppression. The predictions plotted on the right panel show better agreement with the data or even underestimate the suppression. The predictions on the right involve either quark recombination mechanisms \cite{GRANDCHAMP,KOSTYUK,ANDRONIC,BRATKOVSKAYA} or detailed $\jpsi$ transport in medium \cite{ZHU}. Note however that these predictions differ in the way the cold nuclear absorption is accounted for, in the p+p $\jpsi$ production cross-section used for normalization and in the open charm production cross-section entering recombination mechanism.

\begin{figure}[h]
\vspace*{-5mm}
\begin{center}
\includegraphics[width=75mm,clip=true]{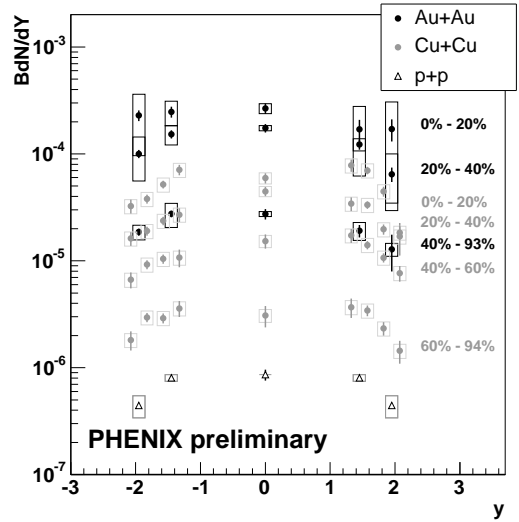}
\end{center}
\vspace*{-15mm}
\caption{\label{fig_y}$\jpsi$ yield as a function of rapidity and centrality.}
\vspace*{-5mm}
\end{figure}

The figure~\ref{fig_y} represents the $\jpsi$ invariant yield $BdN/dy$ as a function of the $\jpsi$ rapidity for different centrality bins in Au+Au, Cu+Cu and p+p collisions. The vertical lines are statistical errors, the bands are point-to-point systematics. Within the large error bars, no significant change in the shape of the distribution is observed from p+p collisions to most central Au+Au. 

\begin{figure}[h]
\vspace*{-5mm}
\begin{center}
\begin{tabular}{cc}
\includegraphics[width=75mm,clip=true]{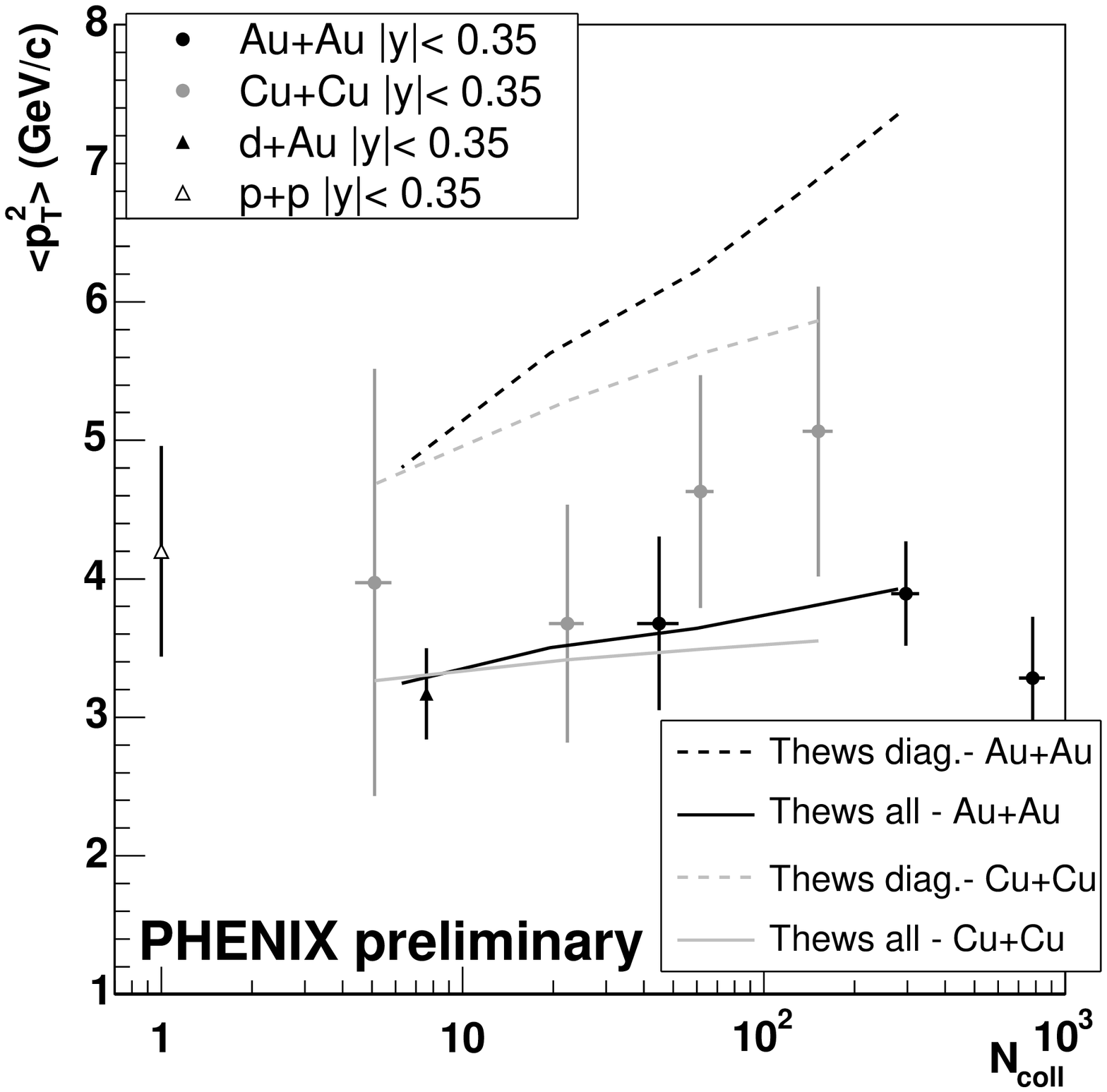}&
\includegraphics[width=75mm,clip=true]{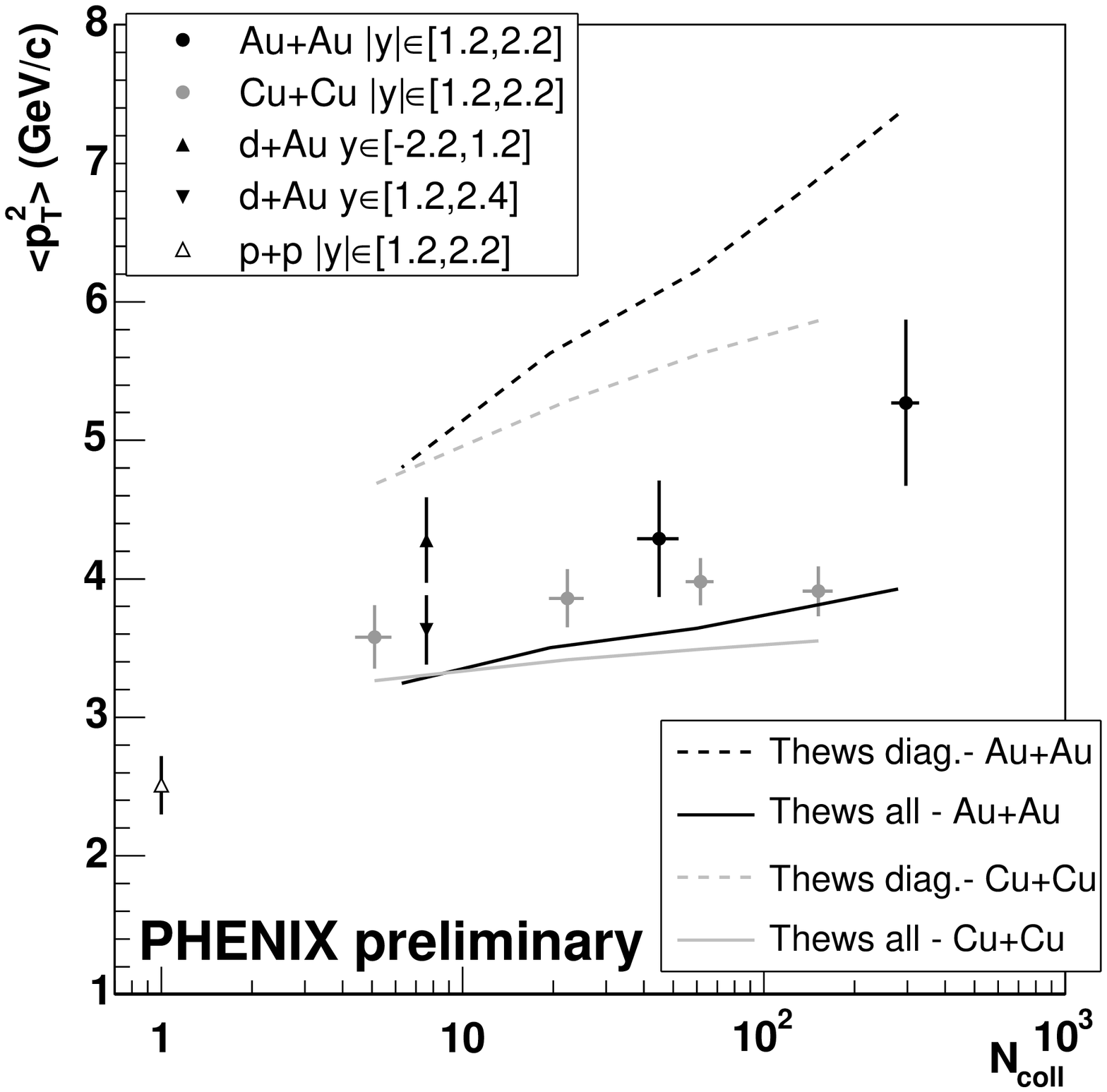}\\
\end{tabular}
\end{center}
\vspace*{-15mm}
\caption{\label{fig_pt}$\jpsi$ mean square transverse momentum as a function of the number of collisions, at mid rapidity (left panel) and forward rapidity (right panel).}
\vspace*{-5mm}
\end{figure}

The figure~\ref{fig_pt} represents the $\jpsi$ mean square transverse momentum as a function of the number of collisions for Au+Au, Cu+Cu, d+Au and p+p collisions. Thews predictions~\cite{THEWS} without (dashed lines) and with recombination (solid lines) are also shown.

\section{Summary}
PHENIX measured $\jpsi$ in Au+Au and Cu+Cu collisions at $\sqrt{s_{NN}}=200$~GeV at both mid and forward rapidity. The dependence of the $\jpsi$ nuclear modification factor on the number of participants allows to discriminate between some of the models capable of reproducing the suppression seen at lower energy. The $\jpsi$ rapidity spectra show no obvious variation in shape. Interpretation of the $\jpsi$ mean square transverse momentum is unclear due to large errors.

\end{document}